\documentstyle[epsf]{mn}

\title[]{{\sl XMM-Newton} observations of polars in low accretion states}

\author[Ramsay et al]{
Gavin Ramsay$^{1}$, Mark Cropper$^{1}$, Kinwah Wu, K. O. Mason$^{1}$, F. A.
C\'{o}rdova$^{2}$ \and W. Priedhorsky$^{3}$\\
$^{1}$Mullard Space Science Laboratory, University College London,
Holmbury St. Mary, Dorking, Surrey, RH5 6NT, UK\\
$^{2}$University of California, Riverside, CA 92521, USA\\
$^{3}$Los Alamos National Laboratory, MS D436, Los Alamos, NM 87545, USA}
\date{Received: }

\begin{document}
\outer\def\gtae {$\buildrel {\lower3pt\hbox{$>$}} \over 
{\lower2pt\hbox{$\sim$}} $}
\outer\def\ltae {$\buildrel {\lower3pt\hbox{$<$}} \over 
{\lower2pt\hbox{$\sim$}} $}
\newcommand{\ergscm} {ergs s$^{-1}$ cm$^{-2}$}
\newcommand{\ergss} {ergs s$^{-1}$}
\newcommand{\ergsd} {ergs s$^{-1}$ $d^{2}_{100}$}
\newcommand{\pcmsq} {cm$^{-2}$}
\newcommand{\ros} {\sl ROSAT}
\newcommand{\exo} {\sl EXOSAT}
\newcommand{\xmm} {\sl XMM-Newton}
\def\rchi{{${\chi}_{\nu}^{2}$}}
\def\uchi{{${\chi}^{2}$}}
\newcommand{\Msun} {$M_{\odot}$}
\newcommand{\Mwd} {$M_{wd}$}
\def\Mdot{\hbox{$\dot M$}}
\def\mdot{\hbox{$\dot m$}}

\maketitle

\begin{abstract}

We have made a series of snap-shot observations of 37 polars using
{\xmm}. We found that 16 of these systems were in a low, or much
reduced, accretion state. Of those, 6 were not detected in
X-rays. This suggests that in any survey of polars, around half will
be in a low accretion state. We tested if there was a bias towards
certain orbital periods: this is not the case. Of the 10 systems which
were detected at low, but significant rates in X-rays, 8 showed
significant variability in their X-ray light curves. This implies that
non-uniform accretion still takes place during low accretion
epochs. The bolometric luminosity of these systems is $\sim10^{30}$
\ergss, two orders of magnitude less than for systems in a high
accretion state. The X-ray spectra show no evidence of a distinct soft
X-ray component. However, the X-ray and UV data imply that such a low
temperature component exists: its temperature is low enough for its
flux distribution to move outside the bandpass of the X-ray
instruments.

\end{abstract}

\begin{keywords}
Stars: individual: -- UZ For, CP Tuc, FH UMa, CV Hyi, AR UMa, V834 Cen, UW
Pic, V393 Pav, V4738 Sgr, RX J0153--59, MN Hya, ST LMi, RX J1313-32, 
V1033 Cen, RS Cae, MR Ser -- Stars: binaries -- 
Stars: cataclysmic variables -- X-rays:
stars
\end{keywords}

\section{Introduction}

Polars or AM Her systems are accreting binary systems in which
material transfers from a dwarf secondary star onto a magnetic
($B\sim$10--200MG) white dwarf through Roche lobe overflow. They vary
in brightness over the binary orbital period, which is synchronised
with the spin period of the white dwarf. Unlike non-magnetic accreting
binaries they do not have a disk and hence a reservoir of
material. Therefore when mass transfer ceases, accretion stops. Many
systems have been found to have episodes in which the accretion rate
is much reduced, or accretion has ceased completely.

The study of Lamb \& Masters (1979) showed that when in a high
accretion state (\mdot \gtae 1 g s$^{-1}$ cm$^{-2}$), the density of
the accreting gas in the accretion column is sufficiently high to
maintain a collisional timescale shorter than the cooling timescale
and the flow dynamical timescale. The flow is hydrodynamic, and an
accretion shock forms near the surface of the accreting white dwarf.
The shock heats the accreting material to temperatures of several tens
of keV and thereby emits keV X-rays (see Wu 2000 and references
therein). There is also a softer X-ray component, due to heating of
the surrounding photosphere of the white dwarf by the hard X-ray
irradiation or by bombardment of dense blobs in the accreting
material.

In contrast, in polars with low accretion rates (\mdot $<$ 0.1 g
s$^{-1}$ cm$^{-2}$) collisions can be so inefficient that not only the
electrons and the ions become thermally decoupled but also ion-ion
collision cannot maintain the flow in hydrodynamic equilibrium. It was
suggested that the kinetic energy of the accreting gas is released via
Coulomb collisions of charged particles in the photosphere of the white
dwarf (Kuijpers \& Pringle 1982; Thomson \& Cawthorne 1987; Woelk \&
Beuermann 1992, 1993). The resulting spectrum would be significantly
different to that found in the high accretion state. This has been
confirmed in X-ray observations of polars seen in high and low
accretion states, see eg the low-state spectra of ST LMi (Ramsay,
Cropper \& Mason 1995) and AM Her (de Martino et al 1998).  In
particular, the soft X-ray component seen in high accretion states is
absent.

However, even during epochs of low accretion rates, the flow is very
unsteady, with flares being seen on short timescales and in various
wavelengths.  In the optical band, AM Her has exhibited flares in the
low state (eg Bonnet-Bidaud et al 2000). Flares are also seen in
X-rays: in the polar UZ For an accretion event was observed using
{\xmm} at the phase when the primary accretion region was in view
(Still \& Mukai 2001, Pandel \& Cordova 2002). The X-ray spectrum was
consistent with a single temperature plasma of temperature $\sim$4keV:
no distinct soft X-ray component was evident.

We have undertaken a survey of 37 polars using {\xmm} (cf Ramsay \&
Cropper 2003 for a preliminary overview of this survey). For details
of those systems observed in a high accretion state, see Ramsay \&
Cropper (2004) and references therein. Our survey also gives us an
indication of how often polars go into a low/off accretion state.
This paper reports on those polars which were in a low state, in
particular their number, light curves and their spectra.

\section{Observations}

{\xmm} was launched in Dec 1999 by the European Space Agency. It has
the largest effective area of any X-ray satellite and also has a 30 cm
optical/UV telescope (the Optical Monitor, OM: Mason et al 2001)
allowing simultaneous X-ray and optical/UV coverage. The EPIC
instruments contain imaging detectors covering the energy range
0.15--10keV with moderate spectra resolution. The OM data were taken
in two UV filters (UVW1: 2400--3400 \AA, UVW2: 1800--2400 \AA) and one
optical band ($V$ band). The observation log is shown in Table
\ref{log}. The data obtained using the RGS instrument (den Herder et
al 2001) were of very low signal-to-noise and are therefore not
discussed further.

The X-ray data were processed using the {\sl XMM-Newton} {\sl Science
Analysis Software} (SAS) v5.3.3.  For the EPIC detectors (Str\"{u}der
et al 2001, Turner et al 2001), data were extracted using an aperture
of 40$^{''}$ centered on the source position. Background data were
extracted from a source free region. The background data were scaled
and subtracted from the source data. The OM data were analysed in a
similar way using {\tt omichain} and {\tt omfchain} in SAS v5.4. Data
were corrected for background subtraction and coincidence losses
(Mason et al 2001). We show in Table \ref{log} the mean $V$ mag
determined using the OM for each source at the time of our
observation.

For a typical polar in a high accretion state (eg V347 Pav, Ramsay et
al 2004), we estimate that the count rate will be a factor of $\sim$4
greater using {\xmm} compared to {\ros}. Every source shown in Table
\ref{log} was found to have a X-ray count rate which was significantly
lower than when it was observed in the {\ros} all-sky survey with the
exception of AR UMa and ST LMi which were not detected and MR Ser when
it was in a low state. However, we extracted two pointed {\sl ROSAT}
observations of AR UMa and found no significant detection in
either. Further, it has been found to show a high X-ray state and an
optical range of $\sim$14.5--16.5 (Schmidt et al 1999) while ST LMi
was found to be bright in X-rays using {\sl EXOSAT} (Beuermann, Stella
\& Krautter 1984). MR Ser was also detected using {\sl EXOSAT}
(Angelini, Osborne \& Stella 1990). Using the highest reported count
rate and an appropriate model, the expected count rate for MR Ser in a
high accretion state using the {\xmm} EPIC pn detector is 4.2 ct/s. We
conclude that all 16 systems in Table \ref{log} were in low or
low/intermediate accretion states.

Of these systems, 10 (UZ For, MN Hya, RX J1313--32, CP Tuc, V834 Cen,
UW Pic, V393 Pav, ST LMi, V1033 Cen and MR Ser) were detected at low,
but significant, levels. We discuss these systems individually in the
next section. We do not discuss UZ For since these {\xmm} data have
been discussed in detail by Still \& Mukai (2001) and Pandel \&
Cordova (2002).

\begin{table*}
\begin{center}
\begin{tabular}{lrlllllrrr}
\hline
Source & Other & Orbit & Date Obs & $V$ & $V$ high       & EPIC 
& RASS & $P_{orb}$ & EPIC MOS\\ 
       & Name    &       &          &     & state          & pn ct/s 
& PSPC ct/s    & (sec) & (sec)\\
\hline
UZ For & & 202 & 2001-01-14 & - & 16(1) & 0.009(1) & 0.70 & 7590 & 19394 \\
CP Tuc & AX J2315--592 & 340 & 2001-10-17 & 19.3 & 17 (2) & 0.018(3) &
0.24 & 5364 & 5395\\
FH UMa & WGA 1047+6335 & 342 & 2001-10-22 & 20.6 & 19 (3) & 0.025(21)
& 0.26  & 4800 & 7792\\
CV Hyi & RX J0132--65 & 362 & 2001-12-01 & 20.6 & 20 (4) & 0.002(2) &
0.28 & 4668 & 7392\\
AR UMa & 1E S1113+432 & 364 & 2001-12-03 & 16.6 & 14.5 (5) & 0.009(8)
& 0.003 & 6954 & 5475\\
V834 Cen & & 403 & 2002-02-19 & 16.0 & 14.3 (6) & hw crash & 7.8\\
         & & 484 & 2002-07-31 & 16.5 & & 0.091(4)$^{*}$ & & 6091 & 6153\\
UW Pic   & RE J0531--46 & 415 & 2002-03-16 & 18.4 & 16.4 (7) &
0.043(5) & 0.73  & 7980 & 6192\\
V393 Pav & RX J1957--57 & 415 & 2002-03-16 & 19.5 & 18 (8)& 0.031(3)&
0.9 & 5929 & 11687\\
V4738 Sgr & RX J2022--39 & 433 & 2002-04-20 & 20 & 18.5 (4) & 0.005(2)
&0.33 & 4680 & 7454\\
RX J0153--59 & & 438 & 2002-04-20 & 19.6 & 17 (9) & 0.005(3) &  0.33 &
4800 & 7494\\
MN Hya & RX J0929-24& 443 & 2002-05-10 & 19.3 & 17.0 (10) & 0.016(3) & 0.33 &
12203 & 7392 \\
ST LMi & & 444 & 2002-05-13 & 17.7 & 15 (11) & 0.029(3)& 0.08 &
6833 & 11975\\
RX J1313-32 & & 484 & 2002-07-31 & 16.3 & 14.6 (12) & 0.31(1) & 1.9 & 15084 & 6884 \\
V1033 Cen & RX J1141--64 & 489 & 2002-08-10 & 20 & 16.5 (13) &
0.044(4)& 0.17 & 11364 & 11286\\
RS Cae & RE J0453--42 & 493 & 2002-08-19 & - & 18.4 (14) & -0.013(7)&
1.3 & 5700 & 9437\\
MR Ser & PG 1550+191 & 571 & 2003-01-22 & 17.3 & 15 (15) & 0.069(6) &
0.04 & 6814 & 9030 \\
\hline
\end{tabular}
\end{center}
\caption{The polars observed in our survey using {\xmm} which were
found to be in low accretion state or `off' accretion state. The
`Orbit' refers to the orbit number of {\xmm} and $V$ was determined
using the {\xmm} Optical Monitor. We show the mean count rate derived
using the EPIC pn with the exception of V834 Cen which was determined
using the EPIC MOS camera. (For sources not clearly detected, we quote
the background subtracted count rate for the expected source
position). The first observation of V834 Cen was affected by hardware
problems. The RASS refers to the {\ros} all-sky survey (for AR UMa we
used PSPC pointed observations, where it was not significantly
detected, and for ST LMi where it was; Ramsay, Cropper \& Mason
1995). The duration of the exposure in the EPIC MOS camera is shown in
the last column.  References for the $V$ high state: (1) Imamura \&
Steinman-Cameron (1998), (2) Thomas \& Reinsch (1996), (3) Singh et al
(1995), (4) Burwitz et al (1997), (5) Schmidt et al (1999), (6)
Imamura, Steinman-Cameron \& Wolff (2000), (7) Reinsch et al (1994),
(8) Thomas et al (1996), (9) Beuermann et al (1999), (10) Sekiguchi,
Nakada \& Bassett 1994 (11) Cropper (1986), (12) Thomas et al (2000),
(13) Motch et al (1996), (14) Burwitz et al (1996), (15) Liebert et al
(1982).}
\label{log}
\end{table*}

\section{Systems showing X-ray variation}

\subsection{V834 Cen}

\begin{figure}
\begin{center}
\setlength{\unitlength}{1cm}
\begin{picture}(8,5)
\put(-0.5,-0.5){\includegraphics{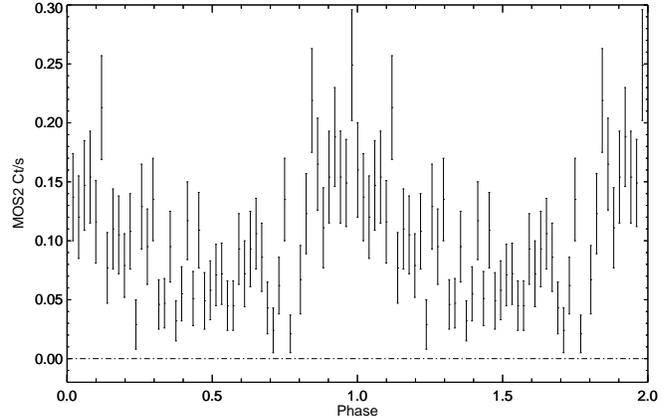}}
\end{picture}
\end{center}
\caption{The phased EPIC MOS2 data (0.15--10.0keV) of V834 Cen phased
on the ephemeris of Cropper, Menzies \& Tapia (1986).}
\label{v834lc}
\end{figure}

Sambruna et al (1994) presented multi-epoch observations of V834 Cen
using {\sl EXOSAT} and found that the shape of its low energy light
curve (0.02--2keV) varied significantly from epoch to epoch. Our EPIC
MOS2 X-ray light curve (Figure \ref{v834lc}) shows significant flux at
all orbital phases and with significant modulation. (Both the EPIC
MOS1 and pn data were in timing mode and are therefore not considered
further). It is most similar to the 1986 {\sl EXOSAT} light curve
(Sambruna et al 1994). However, there is no evidence for an absorption
dip at $\phi$=0.0, even at lower energies, although this maybe simply
due to the relatively poor signal to noise (the accumulated error on
the ephemeris of Cropper, Menzies \& Tapia 1986 is 0.08 cycles).

\begin{figure}
\begin{center}
\setlength{\unitlength}{1cm}
\begin{picture}(8,5)
\put(-0.5,-0.5){\includegraphics{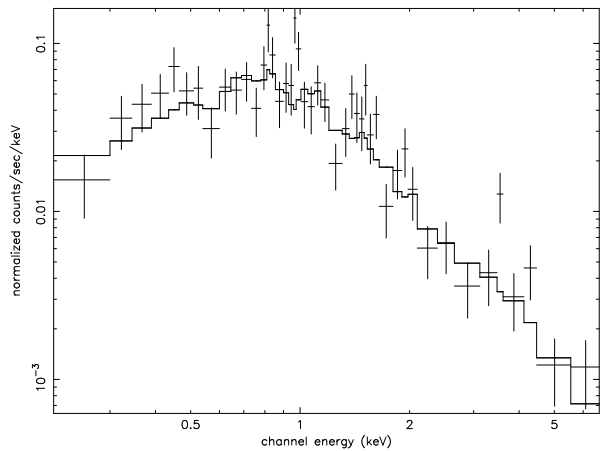}}
\end{picture}
\end{center}
\caption{The EPIC MOS2 spectra of V834 Cen. The best fit is shown as a
solid line and is an absorbed two temperature thermal plasma model.}
\label{v834spec}
\end{figure}

We show in Figure \ref{v834spec} the X-ray spectrum taken using the
EPIC MOS2 detector. We fitted the spectrum firstly using the
stratified accretion column model of Cropper et al (1999). We fixed
the mass of the white dwarf at 0.66\Msun (Ramsay 2000) and the
specific accretion rate at 0.1 g s$^{-1}$ cm$^{-2}$ (appropriate for a
system in a low accretion state). We find a fit with \rchi=1.49 (42
dof). There is no evidence for a distinct soft X-ray component.  

We then fitted the spectrum using a one and two temperature thermal
plasma model (the {\tt MEKAL} model in {\tt XSPEC}). A two temperature
model gave a much better fit than a one temperature fit. The two
temperature model ($kT$=0.3, 4.3keV) is better than the stratified
accretion column model at the 94 percent confidence level. Again there
is no evidence for a distinct soft X-ray component. The bolometric
luminosity ($1\times10^{30}$ \ergsd) is typical of systems in a low
accretion state (=7.4$\times10^{29}$ \ergss for $d$=86 pc, Cropper
1990).

\subsection{UW Pic}

\begin{figure}
\begin{center}
\setlength{\unitlength}{1cm}
\begin{picture}(8,5)
\put(-0.,0.){\includegraphics{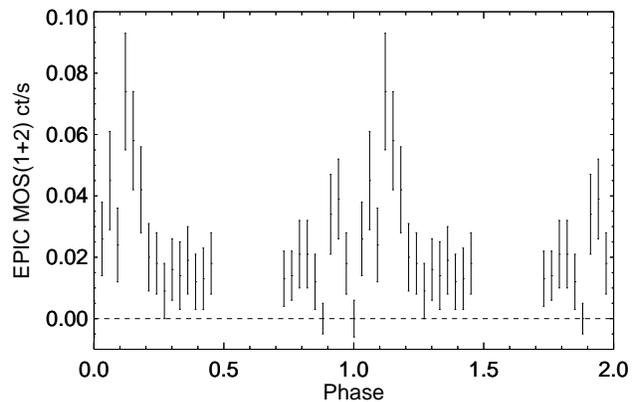}}
\end{picture}
\end{center}
\caption{The phased EPIC MOS1+2 data of UW Pic folded on the period of
Reinsch et al (1994).}
\label{uwpiclc}
\end{figure}

Reinsch et al (1994) present {\ros} survey observations of UW Pic and
show that it has 2 maxima per cycle. Unfortunately their ephemeris is
not precise enough to allow us to phase our data on their phase
convention. However, we do find evidence for a variation in the X-ray
light curve which is similar in shape to the {\ros} light curve,
although the count rate is much lower (Figure \ref{uwpiclc}). The
signal to noise of the integrated spectrum is low. However, if we fit
an absorbed two temperature plasma model ($kT$=0.5, 5keV) to the EPIC
pn spectrum, we find an absorbed, bolometric luminosity of
$1\times10^{30}$ \ergss for a distance of 300 pc (Reinsch et al
1994). We find no evidence for a distinct soft X-ray component.

\subsection{V393 Pav}

\begin{figure}
\begin{center}
\setlength{\unitlength}{1cm}
\begin{picture}(8,5)
\put(-0.,-0.5){\includegraphics{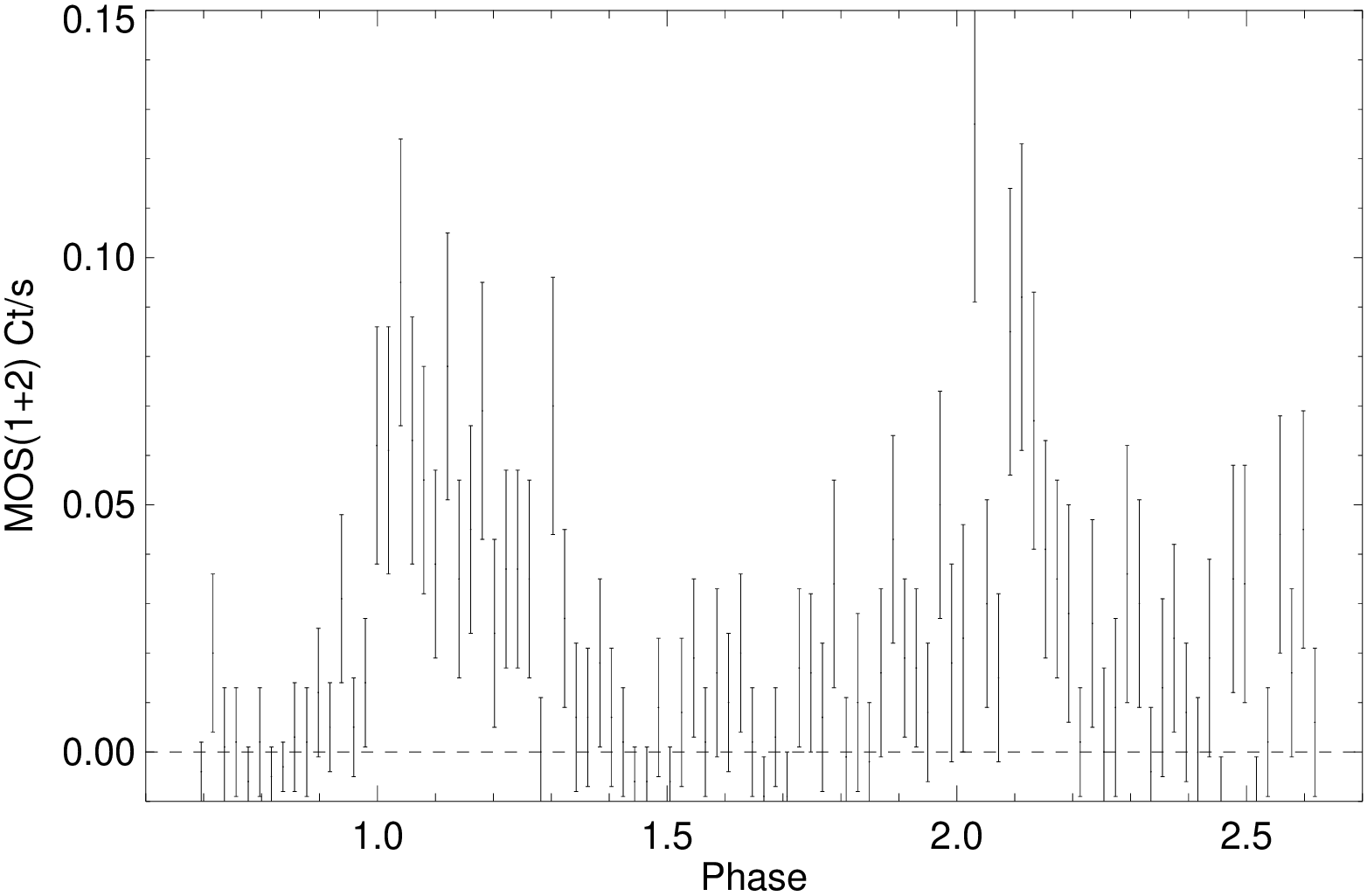}}
\end{picture}
\end{center}
\caption{The EPIC MOS (1+2) data of V393 Pav phased on the ephemeris
of Thomas et al (1996).}
\label{v393lc}
\end{figure}

Optical and X-ray observations reported by Thomas et al (1996) show
V393 Pav to be a one-pole system with a bright phase lasting around
0.5 cycles. Based on the spectroscopic ephemeris the bright phase
occurs at $\phi\sim$0.7-1.2. The accumulated error on the ephemeris up
to the date of the {\xmm} observations is 0.10 cycles. We show in
Figure \ref{v393lc} the 0.15--10.0keV light curve derived using {\xmm}
EPIC pn data.  These data show a variable light curve with significant
flux occurring at $\phi\sim$0.0--0.4, although the bright phase in the
second orbital cycle is more narrow than the first. Although the
signal to noise of the spectra were low, we fitted them with an
absorbed one-temperature thermal plasma model ($kT$=4.5keV) and obtain
a good fit and an unabsorbed bolometric luminosity of
1.3$\times10^{30}$ \ergss assuming a distance of 350 pc (Thomas et al
1996). There was no evidence for a distinct soft X-ray component.

\subsection{ST LMi}
\label{stlmi}

ST LMi was not detected during the {\ros} all sky survey. Further,
when it was the subject of a dedicated observation using {\ros} it was
detected at the low count rate of 0.1 ct/s (Ramsay, Cropper \& Mason
1995). In contrast, the {\sl EXOSAT} observations of ST LMi show the
system to be in a high accretion state and a bright X-ray target ({\sl
ibid}): it shows a bright phase lasting from $\phi\sim$0.7--1.1 on the
ephemeris of Cropper (1986) which is coincident with the optical
bright phase.

\begin{figure}
\begin{center}
\setlength{\unitlength}{1cm}
\begin{picture}(8,5)
\put(-0.,-0.5){\includegraphics{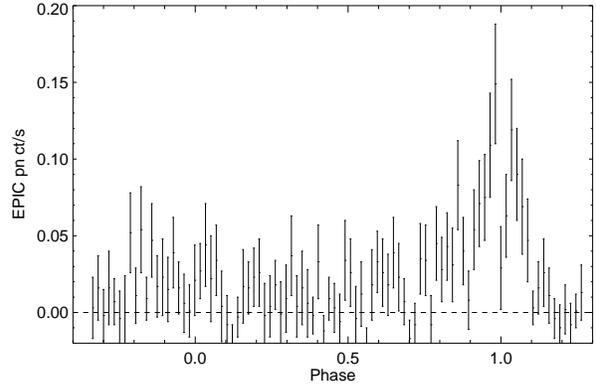}}
\end{picture}
\end{center}
\caption{The phased EPIC pn data of ST LMi folded on the ephemeris
of Cropper (1986).}
\label{stlmilc}
\end{figure}

We show the 0.15--10.0keV light curve made using EPIC pn data in
Figure \ref{stlmilc}. It covers $\sim$1.5 orbital cycles, with the
expected maximum occurring near $\phi\sim$0.0. On the first cycle no
appreciable increase in flux is seen at this phase. However, during
the next cycle there is an increase in flux lasting
$\Delta\phi\sim$0.2 cycles and centered on $\phi\sim$0.0. There is no
corresponding increase in the UVW2 count rate.

\begin{figure}
\begin{center}
\setlength{\unitlength}{1cm}
\begin{picture}(8,4.5)
\put(-0.5,-0.5){\includegraphics{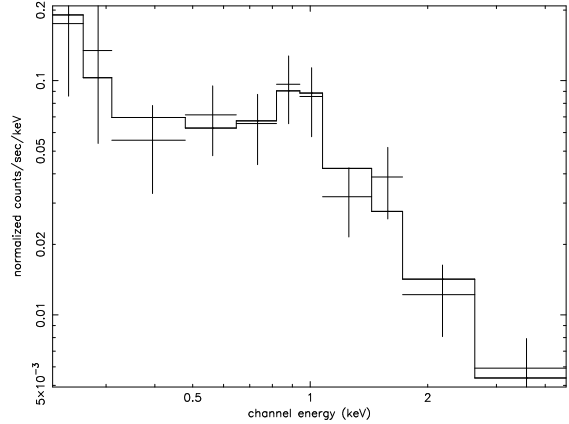}}
\end{picture}
\end{center}
\caption{The EPIC pn spectrum of ST LMi fitted using an absorbed
blackbody plus 2 temperature thermal plasma model.}
\label{stlmispec}
\end{figure}

We extracted a spectrum for the accretion event (which lasted
$\sim$1200 sec). We find it can be well modelled using an absorbed
2-temperature thermal plasma model ($kT$=0.8, 5keV) : no distinct soft
X-ray component is required (Figure \ref{stlmispec}). Using a distance
of 128pc (Cropper 1990) we obtain an unabsorbed, bolometric luminosity
for the accretion event of $1\times10^{30}$ \ergss. Extrapolating this
best fit model to UV energies gives $4.3\times10^{-19}$ \ergscm
\AA\hspace{2mm} in the UVW2 filter. This is many orders of magnitude
less than that observed ($\sim1\times10^{-15}$ \ergscm \AA). This
suggests that the unheated white dwarf contributes a large fraction to
the UV flux.

If we assume that the X-ray event was due to accretion, we can
estimate the mass of the accretion event through,
$L=GM_{1}\dot{M}/R_{1}$, where $M_{1}$ and $R_{1}$ are the mass and
radius of the white dwarf. Assuming $M_{1}$=0.7\Msun and
$R_{1}=8.6\times10^{8}$ cm, $\dot{M}=9.2\times10^{12}$ g s$^{-1}$ or
1.1$\times10^{16}$ g for the duration of the event. This is remarkably
similar to the mass of the accretion event seen in UZ For (Still \&
Mukai 2001), although a corresponding increase in the UVW1 band was
seen in UZ For. This may imply a different temperature for the
reprocessed component in UZ For and ST LMi.

\subsection{RX J1313--32}

\begin{figure}
\begin{center}
\setlength{\unitlength}{1cm}
\begin{picture}(8,4.8)
\put(-0.5,-0.8){\includegraphics{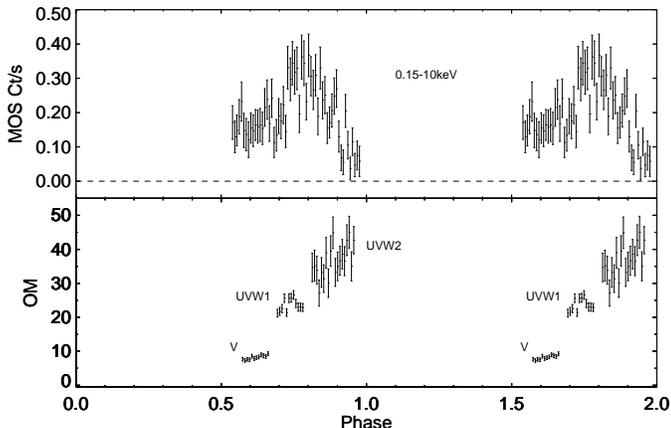}}
\end{picture}
\end{center}
\caption{The phased {\xmm} data of RX J1313--32 folded on the ephemeris
of Thomas et al (2000). Each bin size is 120s. The
units for the OM data are $10^{-16}$ \ergscm \AA$^{-1}$.} 
\label{rx1313lc}
\end{figure}

RX J1313--32 was discovered during the {\ros} all-sky survey when it
was found to be bright reaching a maximum count rate of 6.5 ct/s. When
it was observed again using {\ros} in two pointed observations, it was
found to be much fainter (0.04 ct/s in the PSPC, Thomas et al
2000). Our observations made using {\xmm} show the system reaches a
maximum count rate of 0.4 cts/s (EPIC MOS1+2 combined). Using the
spectral fit described below we find using {\tt PIMMS} (Mukai 1993)
that we would get a count rate in {\ros} of 0.01 and 0.03 ct/s in the
HRI and PSPC respectively. Although the system shows a low count rate
it is still significant and it is likely that RX J1313--32 was
observed in an low/intermediate accretion state (Figure
\ref{rx1313lc}).

\begin{figure}
\begin{center}
\setlength{\unitlength}{1cm}
\begin{picture}(8,5.)
\put(-0.5,-0.5){\includegraphics{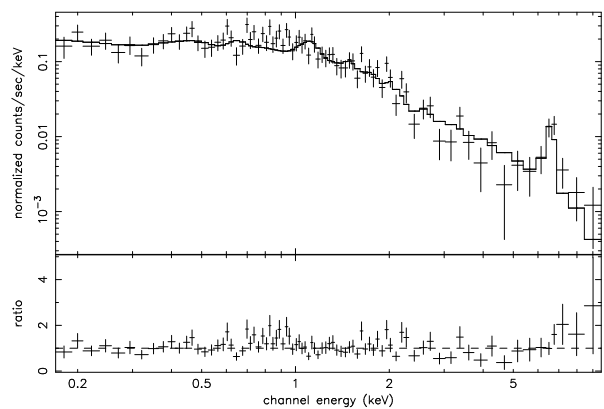}}
\end{picture}
\end{center}
\caption{The integrated EPIC pn spectrum of RX J1313-32 together with
the best model fit.}
\label{rx1313spec}
\end{figure}

We show the spectrum taken using the EPIC pn detector in Figure
\ref{rx1313spec}. There is prominent emission near the Fe K$\alpha$
line. Closer inspection shows that this is due both to a line at
6.7keV and also the fluorescent line at 6.4keV. The spectrum can be
well fit (\rchi=1.04) using a stratified accretion column (Cropper et
al 1999). We assumed a white dwarf mass of 0.7\Msun and 0.1 g s$^{-1}$
cm$^{-2}$. There was no evidence for a distinct soft X-ray
component. The unabsorbed bolometric luminosity is 5$\times10^{30}$
\ergss for a distance of 200pc (Thomas et al 2000).

\subsection{V1033 Cen}

\begin{figure}
\begin{center}
\setlength{\unitlength}{1cm}
\begin{picture}(8,5)
\put(-0.3,-0.5){\includegraphics{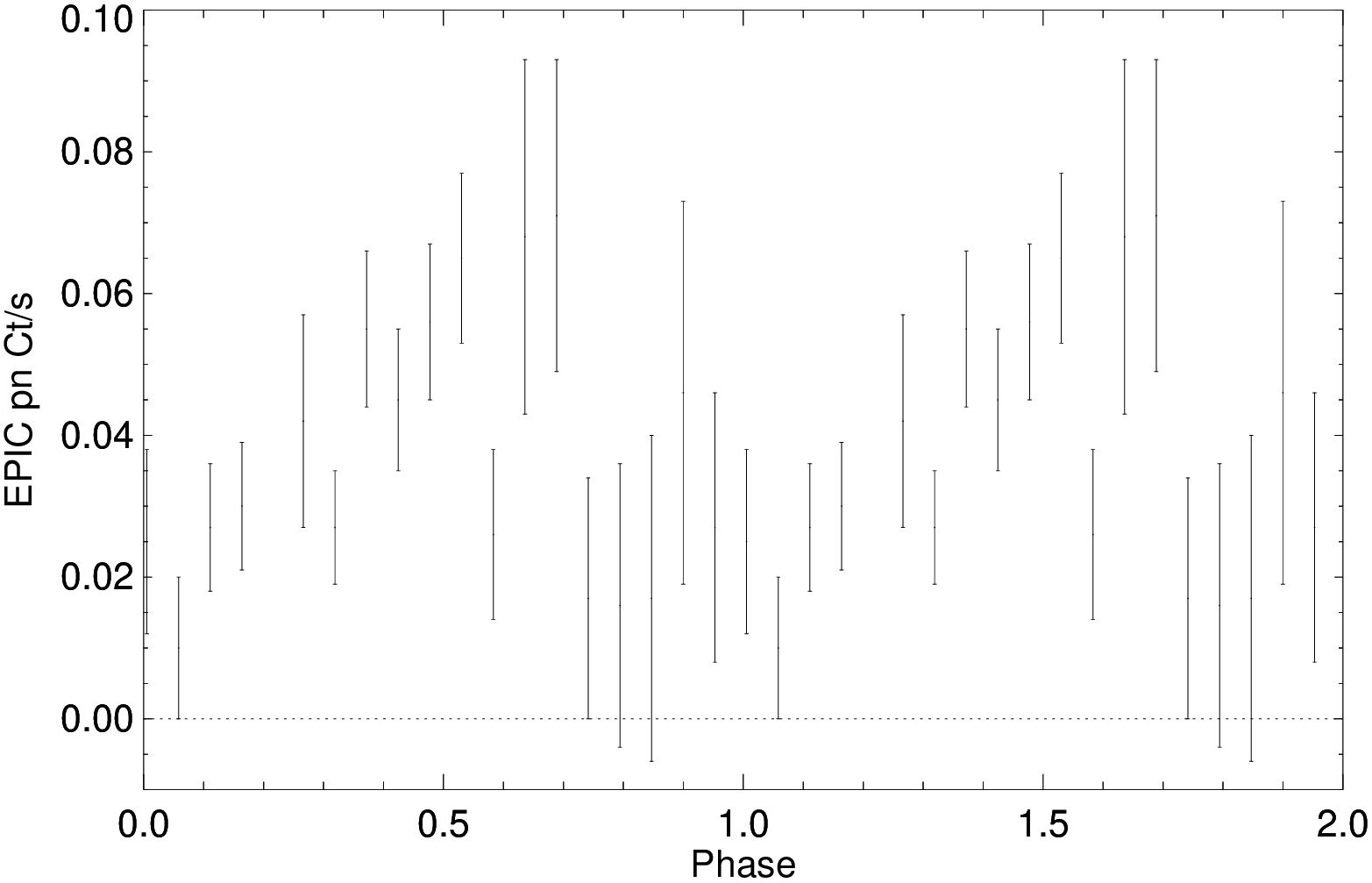}}
\end{picture}
\end{center}
\caption{The phased EPIC pn data of V1033 Cen folded on the ephemeris
of Cieslinski \& Steiner (1997).}
\label{v1033}
\end{figure}

We show in Figure \ref{v1033} the 0.15--10.0keV light curve of V1033
Cen folded on the ephemeris of Cieslinski \& Steiner (1997). It shows
a variation in flux over the orbital period showing maximum near
$\phi\sim0.6$ and a minimum near $\phi\sim$0.7--1.2. However, the
accumulated error in the phasing is 0.37 cycles. Optical photometry
shows a broad maximum around $\phi\sim$0.0 (Buckley et al
2000). Because of the phase error we cannot be certain whether the
X-ray emission originates from this accretion pole.

The spectrum has a low signal-to-noise ratio. However, fitting a
one-temperature thermal plasma model ($kT$=3.5keV) we obtain a
luminosity of 1.8$\times10^{29}$ \ergsd\hspace{1mm} for the integrated
spectrum (the distance to V1033 Cen is currently unknown). There is no
evidence for a distinct soft X-ray component.

\subsection{MR Ser}

\begin{figure}
\begin{center}
\setlength{\unitlength}{1cm}
\begin{picture}(8,4.)
\put(-0.,-0.2){\includegraphics{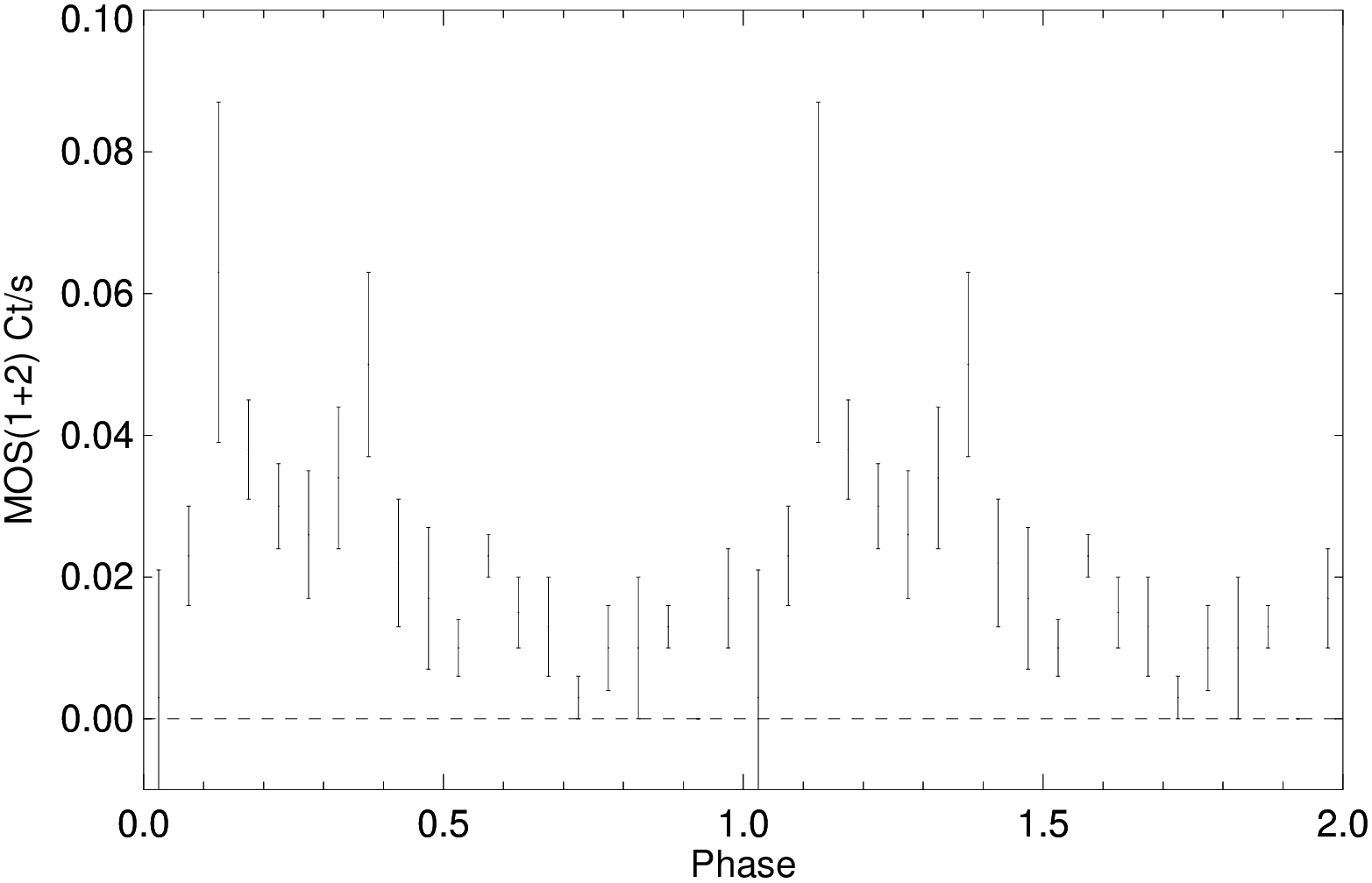}}
\end{picture}
\end{center}
\caption{The phased EPIC MOS(1+2) data of MR Ser folded on the ephemeris
of Schwope et al (1993).}
\label{mrlc}
\end{figure}

MR Ser was discovered using the Palomar Green survey of blue stellar
objects (Liebert et al 1982) and classified as a polar based on its
optical spectrum and variable polarisation. Our observations of MR Ser
show a clear variation in the X-ray flux (Figure \ref{mrlc}). It shows
a positive signal throughout the orbital cycle with a bright phase
lasting $\sim$0.4 cycles. It is reasonably similar to the light curve
obtained using {\exo} on April 1 1984 (Angelini et al 1990). MR Ser
was also found to be in a low accretion state during both the {\ros}
all-sky survey and pointed observations. We know of no observations
since then which have found MR Ser to be in a high state. We can
obtain a good fit to its X-ray spectrum with an absorbed thermal
plasma model with temperature $\sim$1.4keV. The unabsorbed bolometric
luminosity is $\sim3\times10^{29}$ \ergsd, which for $d$=139 pc
(Schwope et al 1993) gives $\sim6\times10^{29}$ \ergss.

\section{Systems showing no X-ray variation}

\subsection{CP Tuc}

Misaki et al (1996) found a new X-ray variable object which they
classified as a polar. It showed a strongly modulated light curve
which was most prominent at lower energies. Our observations show that
it was detected at the 6$\sigma$ level. However, it does not show
evidence for a significant modulation.

\subsection{MN Hya}

MN Hya was discovered during the {\ros} all-sky survey. Pointed {\ros}
observations show a bright X-ray source with an absorption dip just
before the eclipse of the secondary star (Buckley et al 1998). The
{\xmm} observations show a detection at the 5$\sigma$ level. There was
no significant variation in the light curve.

\subsection{The source of X-rays in these systems}

Is the source of X-rays in these systems due to accretion at a very
low rate or is it due to emission from the secondary star? Late type
stars show significant X-ray emission, generally in the form of
flaring activity. We now consider if the X-ray flux from these sources
can be attributed to the secondary star.

Randich et al (1996) found that in late type systems $L_{X}/L_{bol}$
does not exceed $\sim1\times10^{-3}$. We show in Table \ref{latetype}
the secondary mass derived using the standard mass-period relationship
and the maximum X-ray luminosity using the mass-luminosity
relationship of Malkov, Piskunov \& Shipil'kina (1997) assuming a
saturated X-ray flux.

A late type secondary would show an X-ray spectrum dominated by a
thermal plasma of temperature $\sim$1keV. To estimate the expected
count rate, we assume this emission model and set the absorption to be
$1\times10^{20}$ \pcmsq. For the EPIC pn detector and a thin filter,
we find that a count rate of 0.01 ct/s would give a luminosity of
$\sim2\times10^{28}$ \ergsd.

We find that in both CP Tuc and MN Hya the inferred luminosity is
significantly greater than that expected for a late type main sequence
star. We conclude that we have not detected the secondary star in
either system and the flux is probably due to continuing low-level
accretion.

\small
\begin{table}
\begin{tabular}{lrrrrr}
\hline
Source &         & $L_{X}$ &         & Observed  & Observed \\ 
       & $M_{2}$ & ($\times10^{27}$  & D & $(\times10^{28}$ &
$ (\times10^{28}$\\
       & (\Msun) & erg& (pc)    & ergs s$^{-1}$ &
ergs s$^{-1}$)\\
       &         &   s$^{-1}$ )  &         & $d^{2}_{100}$) & \\
\hline
CP Tuc & 0.1 & 5 & $>$175& 3.6(0.6) & 11.1 (175pc)\\
MN Hya & 0.3 & 50 & $\sim$500& 3.2(0.6)& 80 (500pc)\\
\hline
\end{tabular}
\caption{Those systems which were detected in X-rays but found no
X-ray modulation. We show the secondary mass estimated using the
mass-period relationship, and the maximum X-ray luminosity assuming
the mass-luminosity relationship of Malkov, Piskunov \& Shipil'kina
(1997) and a saturated X-ray luminosity of $L_{X}/L_{bol}$ =
$1\times10^{-3}$. We also show the estimated distance and luminosity
using the count rate to flux conversion described in the text.}
\label{latetype}
\end{table}
\normalsize

\section{The UV observations}

Out of the 16 systems which were found to be in a low state using
{\xmm}, 10 were detected in both the UV filters (UVW1 has an effective
wavelength of 2910 \AA\hspace{1mm} and UVW2 2120
\AA\hspace{1mm}). Another 3 systems (FH UMa, CV Hyi and RS Cae) were
detected in the UVW1 filter but not the UVW2 filter. A further 3
systems, (AR UMa, V4738 Sgr and RX J0153--59) were detected in the UV
but not in X-rays. UZ For had data taken in only the UVW1 filter while
V834 Cen had a technical problem with the UVW2 filter observation. We
show in Table \ref{uvflux} the counts in each filter for the 10
systems: we have corrected the observed count rate for coincidence
losses (Mason et al 2001).

As part of the XMM-OM UV calibration process, 8 white dwarfs of
various temperatures have been observed using the OM. These give
conversion factors of count rate to flux (1 ct/s in UVW1 corresponds
to a flux of 4.4$\times10^{-16}$ \ergscm \AA\hspace{1mm} and 1 ct/s in
UVW2 5.8$\times10^{-15}$ \ergscm \AA -- taken from the ESA XMM web
site). These values give fluxes to within 8 percent of their fluxes at
these wavelengths as observed using HST. We show the implied fluxes in
the two filters in Table \ref{uvflux}.

\subsection{Comparing UV count ratios with white dwarf models}

\begin{figure}
\begin{center}
\setlength{\unitlength}{1cm}
\begin{picture}(8,5.5)
\put(-0.9,-0.5){\includegraphics{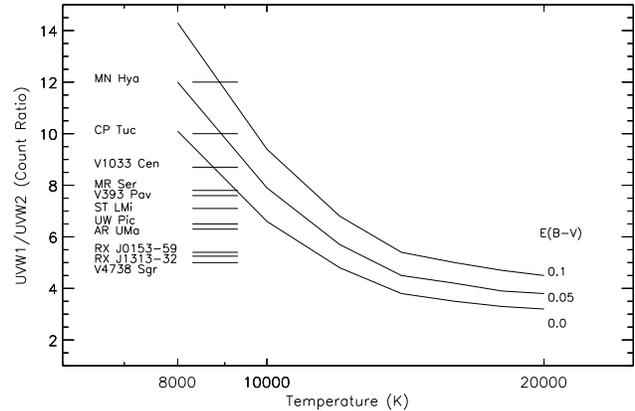}}
\end{picture}
\end{center}
\caption{We show the UVW1/UVW2 count rate ratio for different white
dwarf temperatures and absorption. We have used Hydrogen white dwarf
model atmospheres (those of Koester) and the effective areas of each
UV filter. The count ratio for each low state system is marked at the
left hand side.}
\label{ctratio}
\end{figure}

We show in Figure \ref{ctratio} the ratio of UVW1/UVW2 counts
predicted by convolving white dwarf Hydrogen atmosphere models of
different temperatures (those of Detlev Koester) with the effective
area curves of the UV filters. We predict the ratios for a range in
extinction ($N_{H}=E_{B-V} 4.8\times10^{21}$ \pcmsq, Bohlin, Savage \&
Drake 1978). Polars typically have relatively low levels of
extinction, so we take an `upper-limit' of $E_{B-V}=0.1$.  We used the
extinction law of Seaton (1979).

The UVW1/UVW2 count ratios shown in Table \ref{uvflux} are all greater
than 5. Assuming low extinction this implies temperatures lower than
$\sim$14000K. Those systems with the highest ratios (CP Tuc and MN
Hya) imply temperatures of $\sim$9000K for low extinction. These
temperatures are significantly cooler than the temperatures inferred
from UV spectra of polars in a high accretion state: typically an
unheated white dwarf has $T\sim$15000-25000K together with a hotter
spot (G\"{a}nsicke 1998).

\subsection{Comparing low state and high state systems}

We now compare the UV ratios for systems in the low state with those
polars which were observed in a high accretion state (these are taken
from Ramsay \& Cropper 2004). Table \ref{uvflux} shows that the
UVW1/UVW2 Count ratio for the low state systems is $>$5, with a mean
of 7.4. In contrast the high state systems show a ratio $<$1.2 with a
mean of 0.7. There is a clear difference between the high and low
state systems: at face value this implies that in the high state the
white dwarf is hotter compared with low state.

\begin{figure}
\begin{center}
\setlength{\unitlength}{1cm}
\begin{picture}(8,6.5)
\put(-1.2,-0.5){\includegraphics{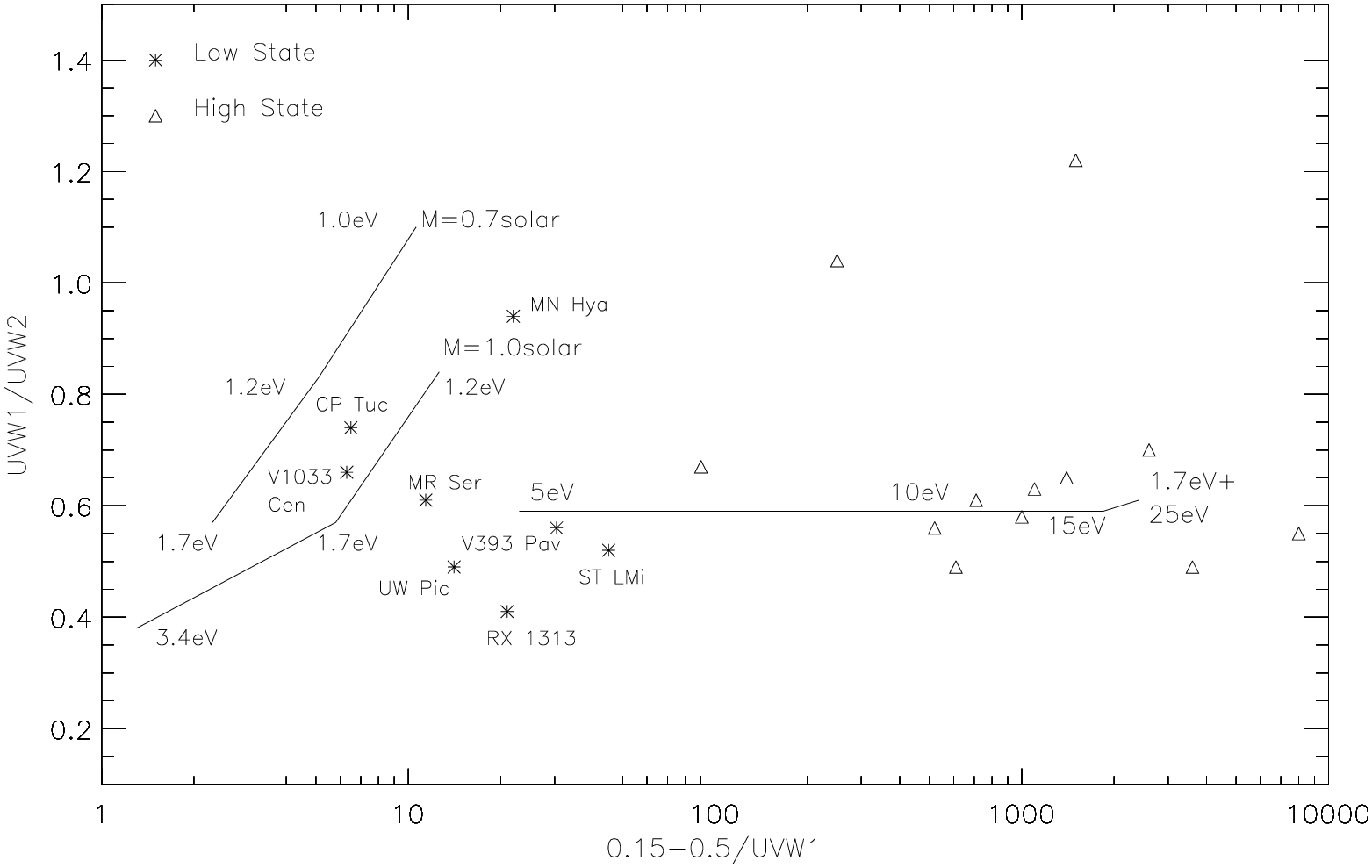}}
\end{picture}
\end{center}
\caption{We show the observed 0.15-0.5keV/UVW1 and UVW1/UVW2 flux
ratios for those systems in a low accretion state and a high accretion
state (taken from Ramsay \& Cropper 2004). For high state systems we
show the expected ratio assuming a polar with mass
0.7\Msun\hspace{1mm} (and using the Nauenberg 1972 relationship) and
temperature 1.7eV (=20000K) and which has a spectrum consistent with
that predicted by the standard accretion shock model (we show the
ratio for various temperatures for the reprocessed soft X-ray
component). We also show the expected ratio for low state systems
assuming a spectrum like that of ST LMi: we also add a blackbody of
various temperature to account for the unheated white dwarf. We assume
the interstellar absorption to be 5$\times10^{19}$ \pcmsq.}
\label{ratio}
\end{figure}

However, in the high state it is common to have a soft X-ray component
which is due to re-processing of hard X-rays from the photosphere of
the white dwarf. This will make some contribution to the UV flux. To
explore this further, we plot the ratios 0.15--0.5keV/UVW1 (flux)
against UVW1/UVW2 (flux) in Figure \ref{ratio} for the systems shown
in Table \ref{uvflux} and also those systems which were observed in a
high accretion state. There is a clear separation of low and high
state systems, with low state systems lying towards the left hand
side, with high state systems in the right hand side.

To determine if these ratios are consistent with that we expect, we
estimate the UVW1/UVW2 and 0.15-0.5/UVW1 flux ratios for both low and
high state systems. For low state systems we take the best fit X-ray
spectral model similar to that of ST LMi (\S \ref{stlmi}): namely a
two-temperature thermal plasma model. We also add a blackbody
component of varying temperature which simulates the unheated white
dwarf (0.7\Msun and 1.0\Msun and assume the Nauenberg 1972 mass-radius
relationship for white dwarfs) and place it at the distance of ST LMi
(128pc Cropper 1990). (Since these estimates are already uncertain and
we are using ratios, we use a blackbody rather than a white dwarf
model atmosphere for the unheated white dwarf as is sufficient for our
purposes). We assume the fraction of the white dwarf which is X-ray
heated is much smaller than total surface area.  For the high state
systems we assume a model consisting of a soft X-ray blackbody
component (of varying temperature) and a thermal bremsstrahlung
component of temperature 30keV. Further, we assume an X-ray flux
distribution consistent with the standard accretion model, ie the soft
X-ray luminosity is approximately half that of the hard X-ray
luminosity (cf Ramsay \& Cropper 2003b). We also include a second
blackbody which simulates the unheated white dwarf: we assume a
temperature of 20000K (=1.7eV; appropriate for the white dwarf in a
high accretion state, G\"{a}nsicke 1998) and a 0.7\Msun white dwarf.
We scale the normalisation of this component so that the soft X-ray
component due from re-processing covers a fraction of 10$^{-4}$ of the
surface of the white dwarf. In both cases we fix the absorption at
5$\times10^{19}$ \pcmsq. For increasing levels of absorption the ratio
0.15--0.5keV/UVW1 will decrease (we include a component for extinction
in the UV based on the reddening law of Seaton 1979).

We plot these tracks in Figure \ref{ratio}. For low state systems
which have no reprocessed X-ray component we expect their ratio's to
lie near the two left hand tracks: eg V1033 Cen, CP Tuc and MN
Hya. Systems which have low temperature reprocessed components will
lie to the lower right of these tracks: eg MR Ser, UW Pic, RX
J1313--32, ST LMi and V393 Pav. Despite the X-ray spectra of the low
state systems not requiring a reprocessed X-ray component, our soft
X-ray and UV data imply that even in the low state, a reprocessed
component exists: it is not detected in X-rays because the temperature
is low enough to have moved to energies lower than that of soft
X-rays.  In this case, the UVW1/UVW2 ratio is insensitive to the
temperature of the unheated white dwarf until temperatures of less
than $\sim$5eV are reached.

\begin{table*}
\begin{tabular}{lllllcr}
\hline
Source & \multicolumn{2}{c} {UVW1} & \multicolumn{2}{c} {UVW2} &
0.15--0.5keV & UVW1/UVW2\\
       & Count& Flux & Count & Flux & Flux & (Counts)\\
      & Rate & $\times10^{-16}$  & Rate & $\times10^{-16}$ &
$\times10^{-15}$ & \\
      &(Ct/s) & \ergscm \AA& (Ct/s)& \ergscm \AA& \ergscm & \\
\hline
CP Tuc & 0.60(3) & 2.6(1) & 0.06(1) & 3.5(2) &1.7 & 10.0\\
AR UMa & 10.08(9) & 44.4(5) & 1.61(5) & 93(2) & - & 6.3\\
UW Pic & 1.62(7) & 7.1(1) & 0.25(2) & 14.5(5)& 10.0 & 6.5\\
V393 Pav & 0.53(3) & 2.3(1) & 0.07(1) & 4.1(6)& 7.0 & 7.6\\
V4738 Sgr & 0.15(3) & 0.7(1) & 0.03(1) & 1.7(6)& - & 5.0\\
RX J0153--59 & 0.42(2) & 1.9(1) & 0.08(1) & 4.7(7)& - & 5.3\\
MN Hya & 0.36(3) & 1.6(1) & 0.03(1) & 1.7(2) & 3.5 & 12.0\\
ST LMi & 1.78(3) & 7.8(1) & 0.25(1) & 1.5(6)& 35 & 7.1\\
RX J1313--32 & 5.81(8) & 26.1(3) & 1.10(3) & 64(1)& 55& 5.3\\
V1033 Cen & 0.61(3) & 2.7(1) & 0.07(1) & 4.1(6)& 1.7 & 8.7\\
MR Ser & 3.11(4) & 14.0(2) & 0.40(2) & 23(1)& 16 & 7.8\\
\hline
\end{tabular}
\caption{For those systems which were detected in both the UVW1 (peak
response 2910 \AA) and UVW2 (21210 \AA) filters we show their count rate
in each filter and the corresponding flux. The number in parenthesis
is the error (standard deviation) on the last figure. We also show the
observed flux in the 0.15--0.5keV energy band.}
\label{uvflux}
\end{table*}

\section{The frequency of systems in a low accretion state}

In our {\xmm} X-ray survey, we observed a sample of 37 systems, which
comprises a significant portion of the currently known polars.  The
orbital periods of this sample range from 77 min to 270 min.  Sixteen
systems (43 percent) of the systems were found in a state of low or
much reduced accretion rate.

Mass transfer in magnetic cataclysmic variables, including polars, is
believed to be governed by the orbital evolution of the binary.  The
frequency of high/low state transitions and the durations of the state
duty cycles determine the observed proportion of high or low-state
systems. We ask: are the high/low state transition and duty cycle of
polars governed by processes dependent on the orbital period?  If so,
association of the high/low states with the orbital periods will be
reflected in our {\xmm} polar survey sample. To verify this we carry
out several statistical tests, (i) to examine the bias of the high and
low-state systems toward long or short orbital periods, (ii) to verify
whether or not there is an alternating sequence of high/low state
transition (including periodic transition) which can introduce
clumping of particle states at certain orbital periods, and (iii) to
determine the relative duration of high/low-state duty cycles.

We note that polars have previously been surveyed using {\ros}. We
therefore also re-analyse the {\ros} sample and compare it with our
{\xmm} sample.  In defining a suitable sample from the {\ros} data, we
consider all polars which were {\sl not} discovered using the {\ros}
all-sky survey. We also exclude the very faint systems which may not
be detected in the relatively short exposure of the {\ros} all-sky
survey (eg V1500 Cyg, EU Cnc which are relatively distant systems).
We performed a similar study to that done for the {\xmm} sample -
namely we compared the count rate for each source determined during
the {\ros} all-sky survey with other X-ray observations of the same
source. Out of a total of 28 systems, 16 (57 percent) were in a low
accretion state during the {\ros} all-sky observation.

In the statistical analysis of the two samples, we assume that the
duration of the high state and low states ($\sim$ weeks, months or
longer) are much longer than exposure time of individual systems
(typically $<$10 ksec) during the survey.  Under this assumption, the
observation of each system can be treated as a snapshot.

\begin{figure}
\begin{center}
\setlength{\unitlength}{1cm}
\begin{picture}(6,6.5)
\put(-1,-1){\includegraphics{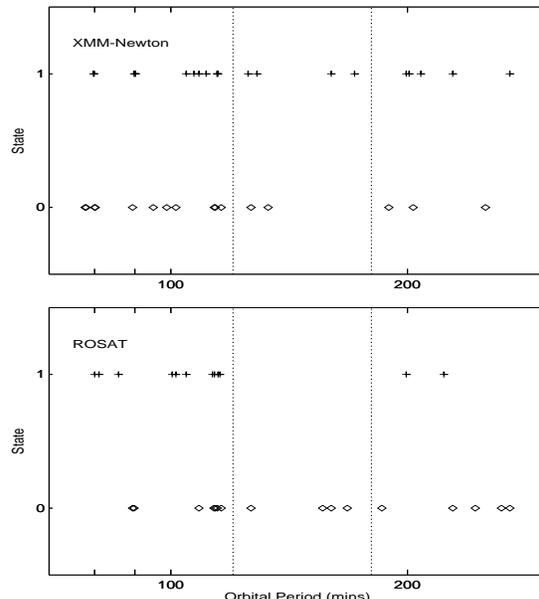}}   
\end{picture}
\end{center}
\caption{The period distribution of polars in high and low accretion
state (assigned with values 1 and 0 respectively) during the {\xmm}
survey (top panel) and the {\ros} all-sky-survey.  We also mark the
periods of 120 and 180 min (dotted vertical lines), the approximate
boundaries of the period gap, the range of periods where magnetic
cataclysmic variables have been claimed to be deficient.}
\label{high_low}
\end{figure}

\subsection{High/low state duty cycle and orbital-period} 

As we do not have information about the functional forms of the
orbital period distributions of the high-state and the low-state
systems, conventional parametric analyses such as those involving the
means and the variances of the two distributions are not directly
applicable.  To overcome this difficulty, we consider non-parametric
tests instead.  These tests can be applied even for situations in
which the numerical scales of the measurements are arbitrary, thus
allowing us to examine how the high/low state transition probability
is related to the orbital period in an objective manner, without
pre-assumed models for the association.
        
We consider a Wilcoxon-Mann-Whitney rank sum U-test (Wilcoxon 1945,
van der Waerden 1969) to search for biases of high and low-state
systems towards long/short orbital periods and a general test of
randomness (see eg Johnson 1994) to examine the possibility of orbital
clumping of high-state and low-state systems.  Both tests require the
systems to be ranked, and we choose a simple criterion: ranking
according to the orbital period.  We also adopt the two following
conventions in our analysis.  (1) The system with the shortest period
has the 1st rank, and a higher rank for longer-period systems.  (2)
Each low-state system is assigned with a state identity parameter of
value L, each high-state system with a state identity parameter of
value H.  The ranks of the systems in the {\xmm} and {\ros} surveys
are shown in Table \ref{rank}, while the period distribution of the
systems are shown in Figure \ref{high_low}.

\begin{table}
\begin{center}
\begin{tabular}{lcrlcr}
\hline
\multicolumn{3}{c}{\xmm} & \multicolumn{3}{c}{\ros}\\
\hline
$P_{orb}$ & H/L? & Source & $P_{orb}$ & H/L? & Source \\ 
\hline
      77.8     & L & J0132-65 &        80.    & H & FH UMa     \\    
      78.0     & L & J2022-39 &        81.02  & H & EF Eri      \\  
      79.7    & H &EV UMa &            85.8   & H & J0425-57 \\ 
      79.9    & H & J1015+09 &       89.4   & L & CP Tuc         \\
      80.0     & L & J0154.0-5947 & 89.8   & L & DP Leo         \\
      80.15    & L & FH UMa &          100.4  & H & VV Pup         \\
      89.04    & L & CP Tuc &          101.5  & H & V834 Cen       \\
      89.8    & H &DP Leo &            104.6  & H & EP Dra         \\
      90.08   & H &V347 Pav &          108.6  & L & CE Gru         \\
      90.14   & H & J1149+28 &       112.97 & H & V2301 Oph      \\
      95.0     & L & J0453-42 &     113.47 & L & MR Ser         \\
      98.8     & L & J1957-57 &     113.64 & H & BL Hyi         \\
      101.5    & L & V834 Cen &        113.9  & L & ST LMi         \\
      104.6   & H &EP Dra &            114.6  & L & WW Hor         \\
      107.0   & H & J1002-19 &       114.84 & H & AN UMa         \\
      108.6   & H &CE Gru &            115.45 & H & EK UMa         \\
      110.9   & H & J2115-58 &       115.92 & L & AR UMa         \\
      113.6    & L & MR Ser &          126.5  & L & UZ For         \\
      113.9    & L & ST LMi &          156.   & L & SDSS1324       \\
      114.84  & H &AN UMa &            160.   & L & V349 Pav       \\
      115.45  & H &EK UMa &            167.7  & L & WX LMi         \\
      115.49  & H &WW Hor &            185.6  & L & AM Her         \\
      115.92   & L & AR UMa &          199.3  & H & BY Cam         \\
      125.4   & H &EU Cnc &            222.5  & H & QQ Vul         \\
      126.5    & L & UZ For &          228.4  & L & VY For         \\
      128.7   & H & J1846+55 &       244.   & L & HS0922+13      \\
      133.0    & L & J0531-46 &     263.4  & L & SDSS1553       \\
      160.0   & H &V2009-65 &          270.   & L & V895 Cen       \\
      171.3   & H & J0501-03 &    & & \\
      189.4    & L & J1141-6410 &   & & \\
      199.3   & H &BY Cam &          & & \\   
      201.0   & H &V1500 Cyg &       & & \\  
      203.4    & L & MN Hya &        & & \\  
      208.0   & H & J1007-20 &      & & \\ 
      228.4   & H &VY For &            & & \\
      251.4    & L & J1313-32 &    & & \\
      270.0   & H &V895 Cen &         & & \\
\hline
\end{tabular}
\end{center}
\caption{The list of sources together with their orbital period and
flagged for whether they were in high or low accretion states at the
epoch of our {\xmm} survey (left) or the {\ros} all-sky survey
(right). For reasons of space we have omitted the `RE' and `RX' prefix
for sources discovered using {\ros}.}
\label{rank}
\end{table}

\subsubsection{Bias in the orbital periods} 
\label{test_1}

First, we verify if there is an excess of either high-state or
low-state systems among the long-period polars and among the
short-period polars.  The null hypothesis is: there is no bias of
high-state and low-state systems towards long orbital periods or short
orbital periods.

In the {\xmm} sample, the number of low-state systems is $n_1 = 16$
and the number of high-state systems is $n_2 = 21$.  The corresponding
rank sums of the two types of systems are $W_1 = 268$ and $W_2 = 435$.
In the Wilcoxon-Mann-Whitney test, the $U$-statistic is given by
\begin{eqnarray} 
  U_i & = & W_i - \frac{n_i \left(n_i+1\right)}{2} 
   \hspace*{0.9cm} (i=1,2)   \nonumber  
\end{eqnarray}    
As the numbers of two types of systems are sufficiently large
($n_1>10$ and $n_2 > 10$), the $U$-statistics are described
approximately by a normal distribution, with the mean and variance
given by
\begin{eqnarray} 
   \mu(U_1) & = \mu(U_2) & = \frac{n_1 n_2}{2} \ , \nonumber \\ 
  \sigma^2(U_1) & = \sigma^2(U_2) &  
       = \frac{n_1 n_2 (n_1 +n_2+1)}{12} \nonumber  
\end{eqnarray}    
The associated standardised random variable is defined as  
\begin{eqnarray} 
  z & = & \frac{U_1 - \mu(U_1)}{\sigma (U_1)} \ .  \nonumber     
\end{eqnarray} 
Thus, the statistics that characterise the {\xmm} sample are $U_1 =
132$, $U_2 = 204$ (cf Table \ref{rank}), $\mu(U_i) = 168$ and
$\sigma^2(U_i) = 1064$, implying a standardised random variable $z_{\rm
xmm} = -1.10$.

In the {\ros} sample, we have $n_1 = 16$ and $n_2 =11$, so the $U$
statistic can again be approximated by a normal distribution.  The
rank sums of the two types of systems are $W_1 = 279$ and $W_2 = 127$;
the relevant statistics are $U_1 = 143$, $U_2 =49$, $\mu(U_i) = 96$
and $\sigma^2(U_i) = 464$.  The corresponding standardised random
variable is $z_{\rm rosat} = +2.18$.
 
For 99 and 95 percent confidence interval, $-2.575 < z < 2.575$ and
$-1.96 < z < 1.96$ respectively.  The $z$ value for the {\xmm} sample
is well within the 95 percent confidence interval, while the {\ros}
sample is within the 99 percent confidence interval. Therefore, we
cannot reject the null hypothesis on a high significance level to
favour the alternative hypothesis.

We note that similar results can be obtained by the Kolmogorov-Smirnov
test, which may be inefficient to discriminate the difference in the
period distribution of the high-state and low-state systems especially
when the difference occurs near the two ends of the period range.  The
$D$-statistic that we obtain for the {\xmm} sample is 0.277,
corresponding to a probability of 0.49 for the null hypothesis.  The
{\ros} sample has a larger $D$-statistics, 0.479, and a
null-hypothesis probability of 0.09.
   
As a summary, we conclude that the {\xmm} sample and the {\ros} sample
do not show significant difference in the (global) distributions of
orbital periods among the high-state and low-state systems between 77
and 270 min.  That is, there is no firm bias of high-state or
low-state systems toward short orbital periods and towards long
orbital periods.
  
\subsubsection{Period clumping of high and low-state systems} 
\label{test_2}
 
We now test whether the systems in the two accretion states are in a
random order when ranked according to their orbital periods.  This
test will reveal the sequence of alternating high/low states in the
population if the alternation depends strongly and explicitly on the
orbital period, but it does not exclude such transitional behaviour
for each individual system.

The null hypothesis is: the period ordering of the high-state and
low-state systems are random.  Using our adopted convention, we
construct a sequence for the {\xmm} sample: LLHHL... (cf Table
\ref{rank}).  We then define a $u$-statistic, which is the number of
runs (see Johnson 1994) in the sequence, and for the {\xmm} sample it
is 20.  For samples with $n_1 > 10$ and $n_2 > 10$, the $u$-statistics
approximately have a normal distribution.  The mean and variance of
the distribution given by
\begin{eqnarray}
\mu(u) & = & \frac{2n_{1}n_{2}}{n_{1}+n_{2}} +1 \ , \nonumber \\ 
\sigma^2(u) & = & 
  \frac{2n_{1}n_{2}(2n_{1}n_{2}-n_{1}-n_{2})}
  {(n_{1}+n_{2})^{2}(n_{1}+n_{2}-1)} \ , \nonumber 
\end{eqnarray} 
  the standardised random variable of the $u$-statistic is  
\begin{eqnarray}  
   z & = & \frac{u - \mu(u)}{\sigma(u)} \ . \nonumber 
\end{eqnarray}  
The statistics of the {\xmm} sample are therefore $u = 20$, $\mu(u) =
19.16$ and $\sigma^2(u) = 8.66$.  It follows that the standardised
random variable $z_{\rm xmm} = 0.29$.
    
Similarly, we can order the {\ros} systems in a similar manner. This
gives $u = 12$, $\mu(u) = 13.71$, $\sigma^2(u) = 6.46$ and $z_{\rm
rosat} = -0.67$.  As $z_{\rm xmm}$ and $z_{\rm rosat}$ is well within
the interval $-1.00 < z < 1.00$, the null hypothesis is consistent
with the data of both survey, implying that the period ordering of the
systems are probably random.  We note that visual inspection may
suggest a deficiency of high-state systems with periods between 120
and 180 min (the controversial period gap of polars) in the {\ros}
sample (Figure \ref{high_low}).  However, this apparent deficiency of
high-state systems is not seen in the {\xmm} sample.  The randomness
test that we have performed indicates that any clumpiness in the two
sequences can be explained by statistical fluctuation.  Combining the
result of this test and that in \S \ref{test_1}, we conclude that the
{\xmm} and {\ros} data do not show evidence that the high-state and
low-state transition and duty cycle of polars are associated with
certain preferred orbital periods.
    
\subsection{Probability of being in the low state} 

Inspection of the proportion of low-state sources suggests that the
fractional duty cycle of low accretion state is 0.5.  However, what is
the uncertainty in this fraction? We show in the Appendix how we
calculate this uncertainty in a formal Bayesian approach.

\begin{figure}
\begin{center}
\setlength{\unitlength}{1cm}
\begin{picture}(8,12.2)
\put(-1.15,5.2){\includegraphics{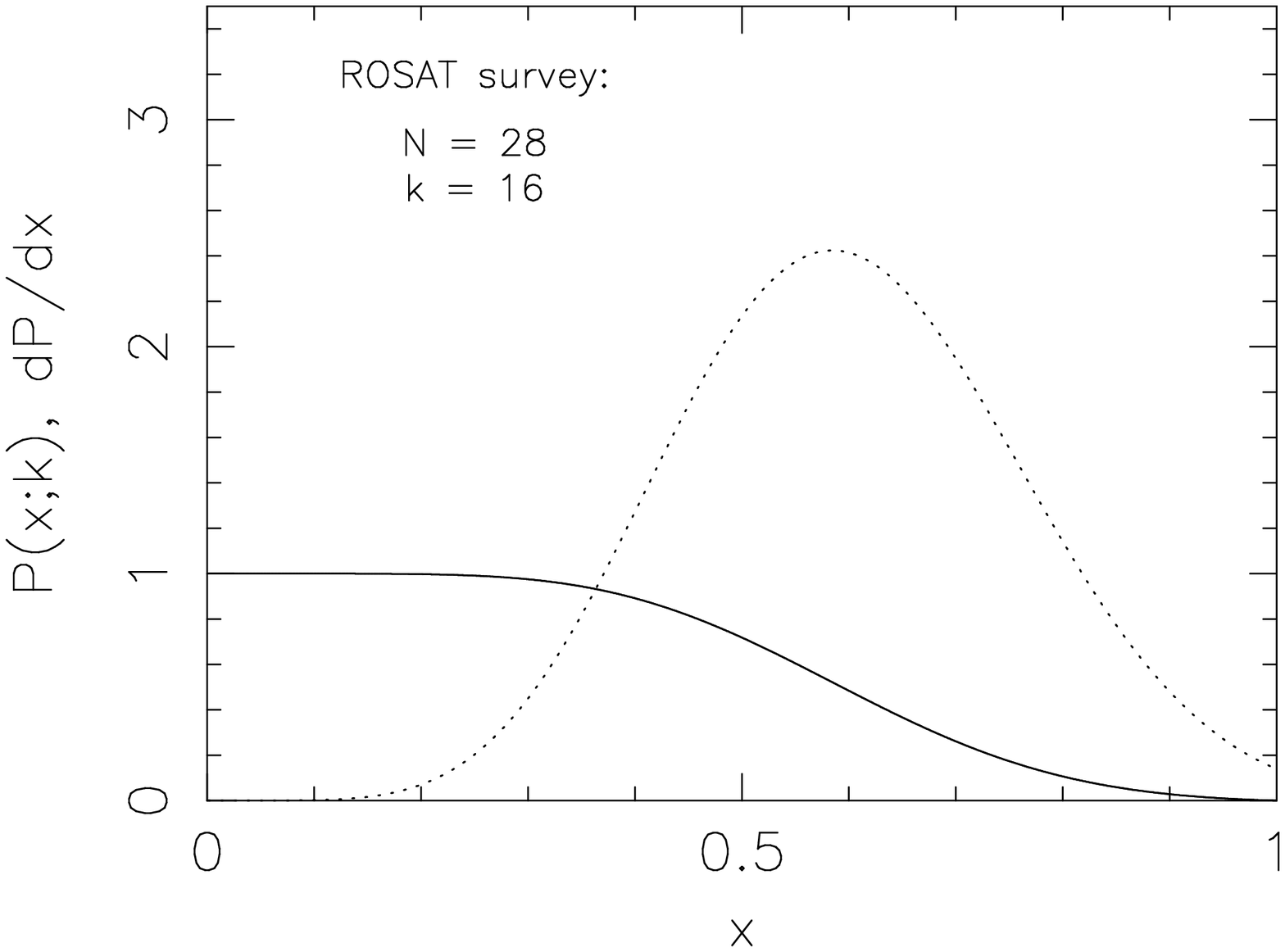}} 
\put(-1.15,-1.10){\includegraphics{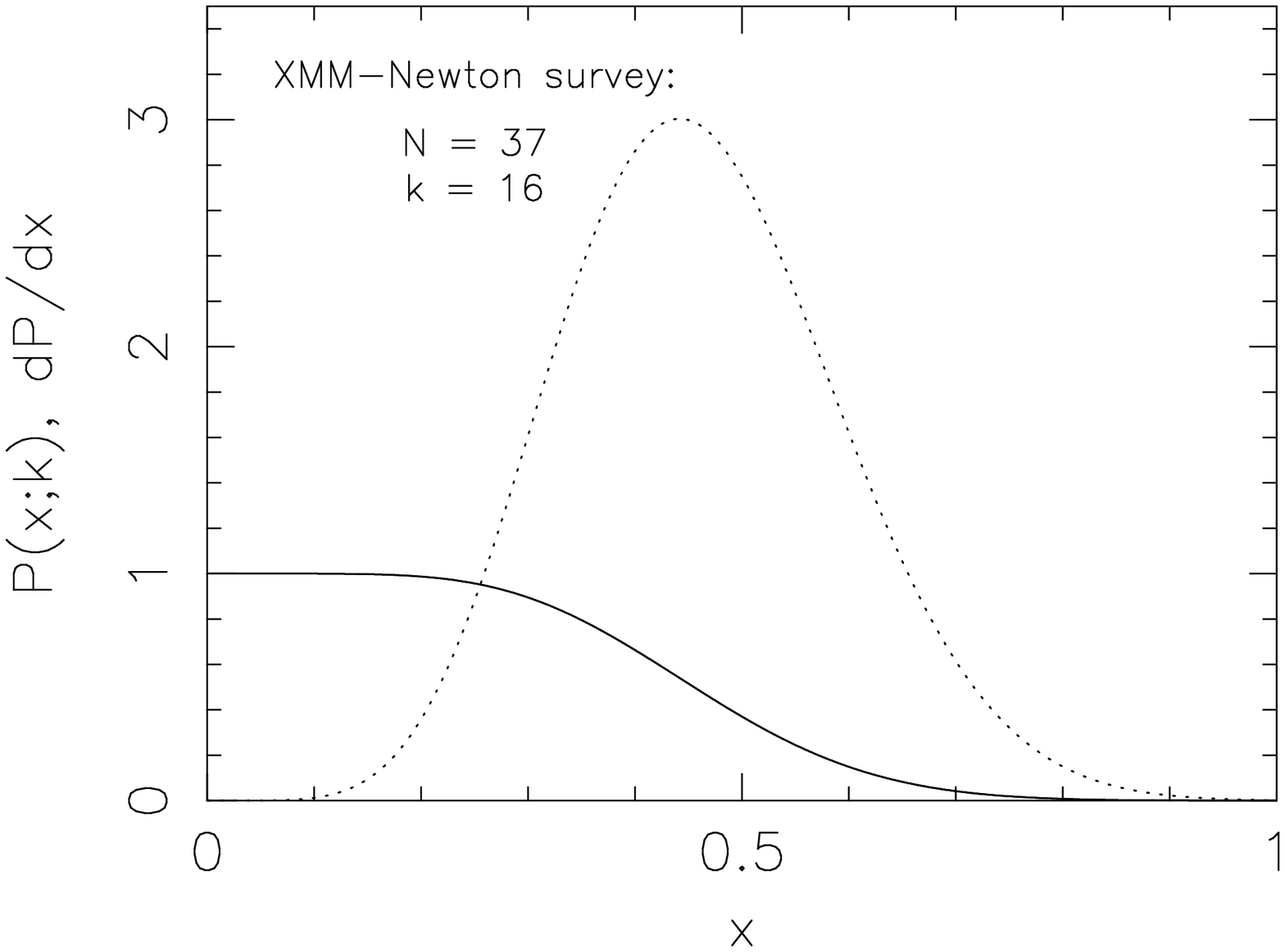}} 
\end{picture}
\end{center}
\caption{The cumulative probability distribution $P(x;k) \equiv
P([x,1]~\vert~k,N)$ (solid line) for the low-state duty cycle $x$ and
its derivative $dP/dx$ (dotted line) for the {\it ROSAT} and {\it
XMM-Newton} surveys (top and bottom panels respectively). The best
estimate of the low-state duty cycle, $\hat x$, is given by the peak
of $dP/dx$, and the bounds of the 90\% confidence interval are located
by $P(x;k) = 0.95$ and 0.05 respectively.}
\label{max_like}
\end{figure}

In the {\it XMM-Newton} sample $N = 37$ and $k=16$, and in the {\it
ROSAT} sample $N=28$ and $k=16$, where $N$ is the number of systems in
the sample and $k$ is the number of systems in a low accretion state.
Figure \ref{max_like} shows the cumulative probability distribution
for low-state duty cycle $x$, which is $P([x,1] | k,N)$, and its
derivative for the two surveys.  The estimate $\hat x$ is 0.442 and
the 90\% confidence interval is $(0.258,0.734)$ for the {\it
XMM-Newton} sample.  The estimate $\hat x$ is 0.585 and the 90\%
confidence interval is $(0.342,0.863)$ for the {\it ROSAT} sample,
consistent with the result obtained from the {\it XMM-Newton}
sample. Thus, we may conclude that the probability of a system in a
low state is $\sim$50 percent, with an uncertainty corresponding to
the two samples given above.

\subsection{Implications for the space density of polars}

Three new polars were discovered using {\exo} in a coverage of 5
percent of the sky. Based on this finding Hameury, King \& Lasota
(1990) predicted that around 60 new polars would be discovered using
{\ros}. Since {\ros} was more sensitive than {\exo}), this number
would be expected to be greater. In contrast, $\sim$30 polars were
discovered using {\ros}.  However, the Poisson uncertainty of the
prediction is $\pm35$.  Taking into account of statistical
fluctuations, we conclude that the detection of approximately 30
systems by {\ros} is consistent with the results of the {\exo} survey.

The estimated space density of CVs is $\sim (1 - 3)\times
10^{-5}$~pc$^{-3}$ (Patterson 1998, Schwope et al 2002a).  The polar to
disk CV ratio, inferred from the {\it Einstein} galactic plane survey,
is about 1/3 (Hertz et al 1990).  Thus, the space density of polars is
probably slightly less than $10^{-5}$~pc$^{-3}$.

Because of inhomogeneities in the local interstellar medium, soft
X-rays sources have very different discovery probabilities in
different parts of the sky.  Soft X-ray sources such as polars and
single white dwarfs, could be preferentially found in certain parts of
the sky (eg Warwick et al 1993).  Thus, the survey carried out by the
{\sl EXOSAT} LE inevitably suffered from this bias.

In addition to the bias due to absorption by an inhomogeneous local
interstellar medium, there are several other limiting factors that
could also introduce uncertainties in the estimates of the space
density of galactic polars.  Among them is the lack of accurate
distance estimates for many systems (many polars have only lower
limits), and the fact that those polars which accrete at an extremely
low rate (cf Schwope et al 2002b) are unlikely to be discovered in
X-ray surveys. Another important limiting factor is the uncertainties
in the proportion of high-state and low-state systems in the
population.  In deriving the space density for polars, many workers
(eg, Thomas \& Beuermann 1998) first assumed that the sample is
complete at high galactic latitudes, then corrected for parts of the
sky (such as the North-Polar Spur) which have higher interstellar
absorption and finally rescaled according to the fraction of systems
in the high-accretion states.  This survey has eliminated one
uncertainty by showing that about a half of the polar population is in
a low accretion state at a given time.

\section{Conclusions}

In our survey of 37 polars using {\xmm} we found that 16 were a low
accretion state.  Of these, 6 were not detected in X-rays.  The {\xmm}
data show no evidence that the low state systems are biased towards
certain orbital periods. The period ordering of the system is
consistent with random. Similar results were found when re-examining a
sample of the {\ros} all-sky survey data, in which 16 out of 28
systems were in low accretion states.

Of those low state systems which were detected at a significant level,
8 showed a significant variation in X-rays. We find that in epochs
with low accretion states, accretion is still occurring, often
sporadically at the same accretion pole as when the system is in a
high accretion state. Their spectra can be modelled using a
two-temperature thermal plasma model of temperature several keV. The
unabsorbed bolometric luminosity is typically $\sim10^{30}$ \ergss:
this is two orders of magnitude less than the luminosity observed in
the high accretion state (Ramsay \& Cropper 2003b). There is no
evidence for a distinct soft X-ray luminosity as often seen in the
high accretion state.  However, the UV data suggests that a
reprocessed component is still present in the low accretion state, but
is cooler than in the high state and therefore shifted into the UV or
far-UV.

\section{acknowledgments}

This is work based on observations obtained with XMM-Newton, an ESA
science mission with instruments and contributions directly funded by
ESA Member States and the USA (NASA). These observations were part of
the OM guaranteed time programme. We thank Detlev Koester for kindly
supplying white dwarf model atmosphere spectra.

\subsection*{APPENDIX: Determining the probability of being in the low state}
\label{appendixa}

The probability of $m$ systems in a low state in a sample of $N$
  system is given by a binomial distribution
\begin{eqnarray} 
   b(m,N;x) & = & \frac{N!}{m!(N-m)!}\ x^m (1-x)^{N-m} \ ,  \nonumber 
\end{eqnarray} 
   where $x$ is the low-state duty cycle of the systems. 
If the systems are not observed simultaneous, 
  it introduces counting fluctuations, 
  which could be approximated by a Poisson process. 
The probability of $k$ systems in a sample of $N$ systems 
  are observed in the low state during a survey is therefore  
\begin{eqnarray}  
   P(k,N~\vert~ x) & = & \sum_{m=1}^{N} \frac{e^{-m} m^k}{k!}
       \ b(m,N;x) \nonumber \\ 
            & = & \sum_{m=1}^{N} \frac{e^{-m} m^k}{k!} 
            \frac{N!}{m!(N-m)!}\ x^m (1-x)^{N-m} \ . \nonumber 
\end{eqnarray} 

From the Bayesian Theorem, 
  the probability of a system being in a low state 
  given that $k$ systems are observed in a low state is 
\begin{eqnarray} 
     P(a\le x\le b ~\vert~ k,N) & = & 
    \frac{\int_a^b dx\ P(k,N~\vert~ x) f(x)}
    {\int^0_1 dx\ P(k,N~\vert~ x) f(x)} \ ,   \nonumber  
\end{eqnarray} 
   where $f(x)$ is a priori probability density distribution of $x$. 
For a uniform density distribution $f(x)$, we have 
\begin{eqnarray}  
    P(a\le x\le b ~\vert~ k,N) & = & \frac{I(a,b;k,N)}{I(0,1;k,N)}  \ ,   \nonumber 
\end{eqnarray}   
  and $x$ can be interpreted as the low-state duty cycle. 
The numerator $I(a,b;k,N)$ is given by 
\begin{eqnarray}  
   I(a,b;k,N) & = &  \sum_{m=1}^{N} \frac{e^{-m} m^k}{k!} 
            \frac{N!}{m!(N-m)!}\ \nonumber \\ 
     & & \hspace{0.5cm}\times \int^b_a dx \ x^m (1-x)^{N-m}   \nonumber \\  
    &   & \hspace*{-2cm} =  \sum_{m=1}^{N} \frac{e^{-m} m^k}{k!} 
            \frac{N!}{m!(N-m)!}\ \nonumber \\ 
   & & \hspace*{-1.75cm} \times \bigg[B_b(m+1,N-m+1) - B_a(m+1,N-m+1) \bigg] 
      \ , \nonumber 
\end{eqnarray}   
   where $B_b(m+1,N-m+1)$ and $B_a(m+1,N-m+1)$ are incomplete Beta functions.  
The denominator $I(0,1;k,N)$ is given by 
\begin{eqnarray}
   I(0,1;k,N) & = & \sum_{m=1}^{N} \frac{e^{-m} m^k}{k!} 
            \frac{N!}{m!(N-m)!}\ \nonumber \\ 
   & & \hspace{0.5cm}\ \times \int^1_0 dx \ x^m (1-x)^{N-m}  \nonumber \\ 
               &   & \hspace*{-2cm}= \sum_{m=1}^{N} \frac{e^{-m} m^k}{k!} 
            \frac{N!}{m!(N-m)!} \  
    \frac{\Gamma (m+1)\Gamma(N-m+1)}{\Gamma(N+2)}  \nonumber  \\ 
            &   & \hspace*{-2cm}= \frac{1}{(N+1)}\ \sum_{m=1}^{N} 
           \frac{e^{-m} m^k}{k!}  \ , \nonumber
\end{eqnarray} 
   where $\Gamma(m+1)$, $\Gamma(N-m+1)$ and $\Gamma(N+2)$ 
     are Gamma functions.

\end{document}